\documentclass[epjCONF,columns]{svjour} 
\usepackage{graphics}
\usepackage[varg]{txfonts} % Times fonts
\usepackage[latin1]{inputenc}
\session-title{Exotic Physics Searches at CMS}
\begin{document}
\title{Exotic Physics Searches at CMS}
\author{Bryan Dahmes on behalf of the CMS Collaboration}
\institute{University of Minnesota, School of Physics and Astronomy, 116 Church Street S.E., Minneapolis, MN 55455}
\abstract{
We summarize the results of several searches for evidence of new physics phenomena using proton-proton collisions 
at $\sqrt{s} = 7$~TeV delivered by the Large Hadron Collider at CERN and recorded by the CMS detector in 2011.
} %end of abstract
\maketitle
\section*{Introduction}
\label{intro}
The Standard Model (SM) has proven itself over many decades to be in very good agreement 
with experimental results.  Despite this success, we realize that the SM is an incomplete 
theory and we therefore anticipate that the data delivered 
by the Large Hadron Collider (LHC) at CERN will contain evidence for new phenomena 
outside the current framework.

In this Note we summarize several searches for exotic new physics processes not 
currently explained by the Standard Model~\footnote{This note is not intended to 
summarize all searches for new physics phenomena conducted by CMS.  Discussions 
of CMS searches for the Higgs boson, Supersymmetry, and exotic resonances can be 
found in other proceedings for this conference.}.
These searches were conducted using 
$\sqrt{s} = 7$~TeV $pp$ collision data delivered by the LHC and collected by the 
CMS detector~\cite{CMS} in 2011.  

\section{Searches for New Heavy Charged Gauge Bosons}
\label{sec:wprime}

Various extensions to the Standard Model predict that new heavy gauge bosons may be found 
with TeV-scale masses.  One such example is the $W'$, which may be considered a heavy analogue 
to the SM W boson and possess the same left-handed fermionic couplings.  CMS searches for 
this boson using 1.1~fb$^{-1}$ of collision data 
via its leptonic decay mode $W' \rightarrow \ell \nu$, reconstructing final states 
containing an electron or muon and a neutrino~\cite{wprime}.  By examining the 
transverse mass distribution of the lepton and neutrino, where the neutrino presence in the 
detector is inferred from the missing transverse energy in the event ($E_T^{\rm miss}$), we 
find the data are in good agreement with SM expectations.  Assuming SM-like couplings, we 
exclude $W'$ bosons with mass below 2.27~TeV at the 95\% confidence level (C.L.).

\begin{figure}[htb]
\begin{center}
\resizebox{0.85\columnwidth}{!}{%
\includegraphics{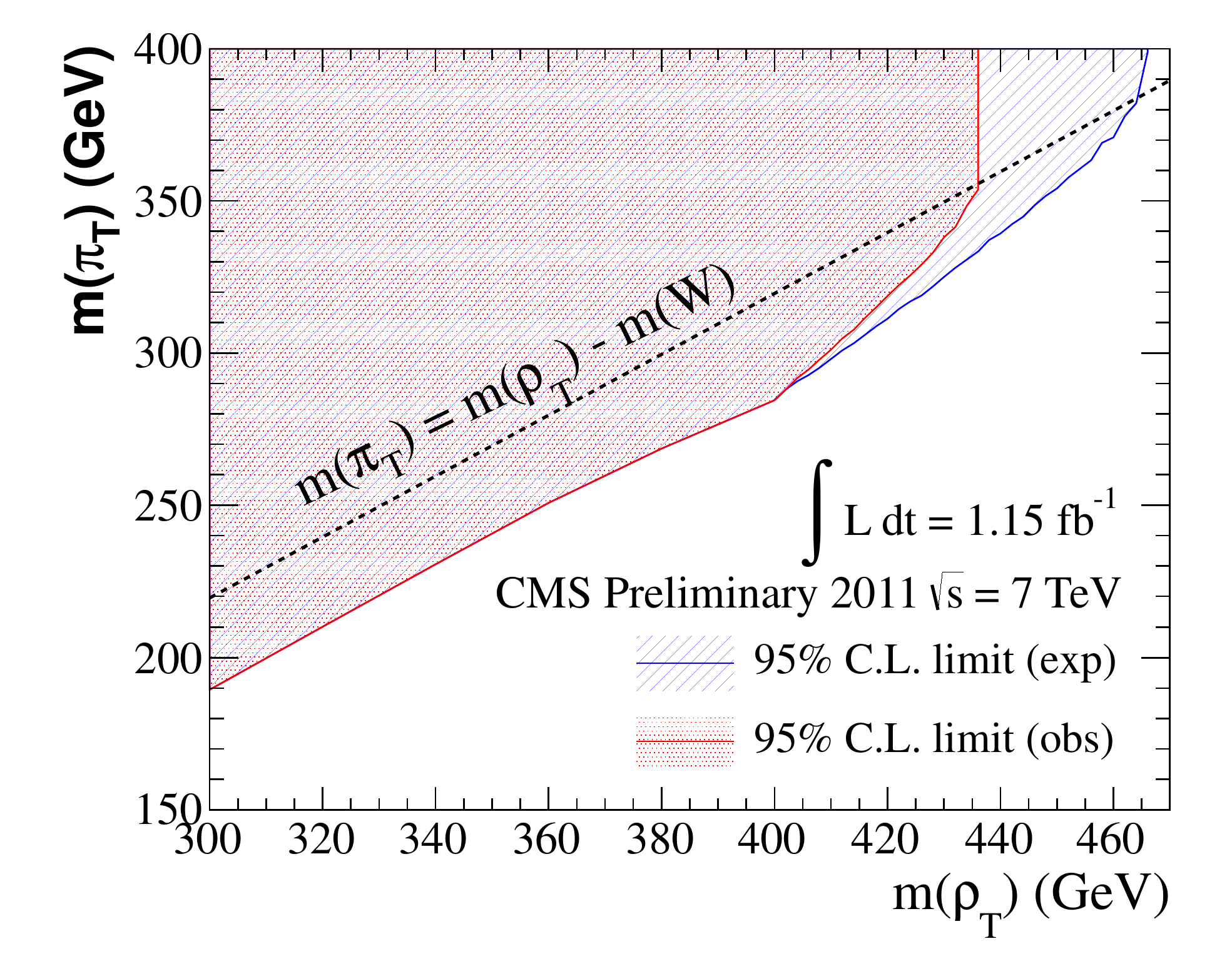} }
\caption{Two-dimensional exclusion limit for Technicolor in the $WZ$ final state 
as a function of the $\rho_{\rm TC}$ and $\pi_{\rm TC}$ masses.}
\label{fig:wprime}     
\end{center}
\end{figure}

Due to the large expected mass of the $W'$, the decay $W' \rightarrow WZ$ is also allowed, 
and if the leptonic decays of the $W'$ are suppressed the decay rate to the $WZ$ 
final state could be relatively enhanced.  We also consider that certain versions of 
Technicolor, a strongly interacting gauge theory which allows for the dynamical 
breakdown of electroweak symmetry, describe techni-particles (e.g. $\rho_{\rm TC}$) that 
may decay to the $WZ$ final state with narrow resonant peaks.
Searching in the $WZ$ final state with 1.1~fb$^{-1}$ of data, where both the 
$W$ and $Z$ decay leptonically, 
we exclude SM-like $W'$ bosons with mass below 784~GeV~\cite{wz}.  We also compare 
our results in the $WZ$ final states to predictions from 
Technicolor models and present 95\% C.L. exclusion for $\rho_{\rm TC}$ 
and $\pi_{\rm TC}$ in Figure~\ref{fig:wprime}.  

Finally, we consider the possibility that the heavy W boson originates from
an extension to the SM that includes a right-handed SU(2) symmetry group 
and search for the decay $W_R \rightarrow \ell N_R$, 
where the heavy neutrino $N_R$ is the right-handed analog of the SM neutrino $\nu_L$.
Assuming right-handed couplings 
for the new heavy gauge boson, using 0.2~fb$^{-1}$ we exclude in the 
two-dimensional $(M_{W_R},M_{N_R})$ space that extends to $M_{W_R} = 1.7$~TeV~\cite{wr}.

\section{Searches for Fourth Generation and Excited Quarks}

Precision electroweak measurements imply that the Standard Model can only accommodate three fermion 
generations.  However, recently there has been renewed interest in extending the SM to include a 
fourth generation of fermions.  Assuming a fourth generation, the indirect bounds on the mass 
of the Higgs boson could be relaxed, and a heavy fourth generation neutrino could be considered as a 
dark matter candidate.  An additional generation of quarks may also possess enough 
intrinsic matter and anti-matter asymmetry to contribute to the baryon asymmetry of 
the Universe.

Given this motivation, CMS searches for fourth generation $b'$ and $t'$ quarks, 
produced singly or in pairs, considering 
multiple final states.  We search for pair-production of $b'$ quarks via the 
decay chain $b' \bar{b}' \rightarrow t W^{-} \bar{t} W^{+}$ and consider final 
states with three leptons ($\ell = e, \mu$) or same-sign dileptons~\cite{bprimePP}.  
We also search for $t' \bar{t}' \rightarrow b W^{+} \bar{b} W^{-}$, and consider final
states where one~\cite{tprime2lep} or both~\cite{tprimeljets} of the $W$ bosons decays 
leptonically.  Standard Model expectations, which are dominated by $t\bar{t}$
production, are in good agreement with the results obtained using 1.1~fb$^{-1}$ of data.
As a result, we exclude heavy top-like (bottom-like) quarks with masses below 
450~GeV (495~GeV) at the 95\% confidence level.

\begin{figure}[htb]
\begin{center}
\resizebox{0.85\columnwidth}{!}{%
\includegraphics{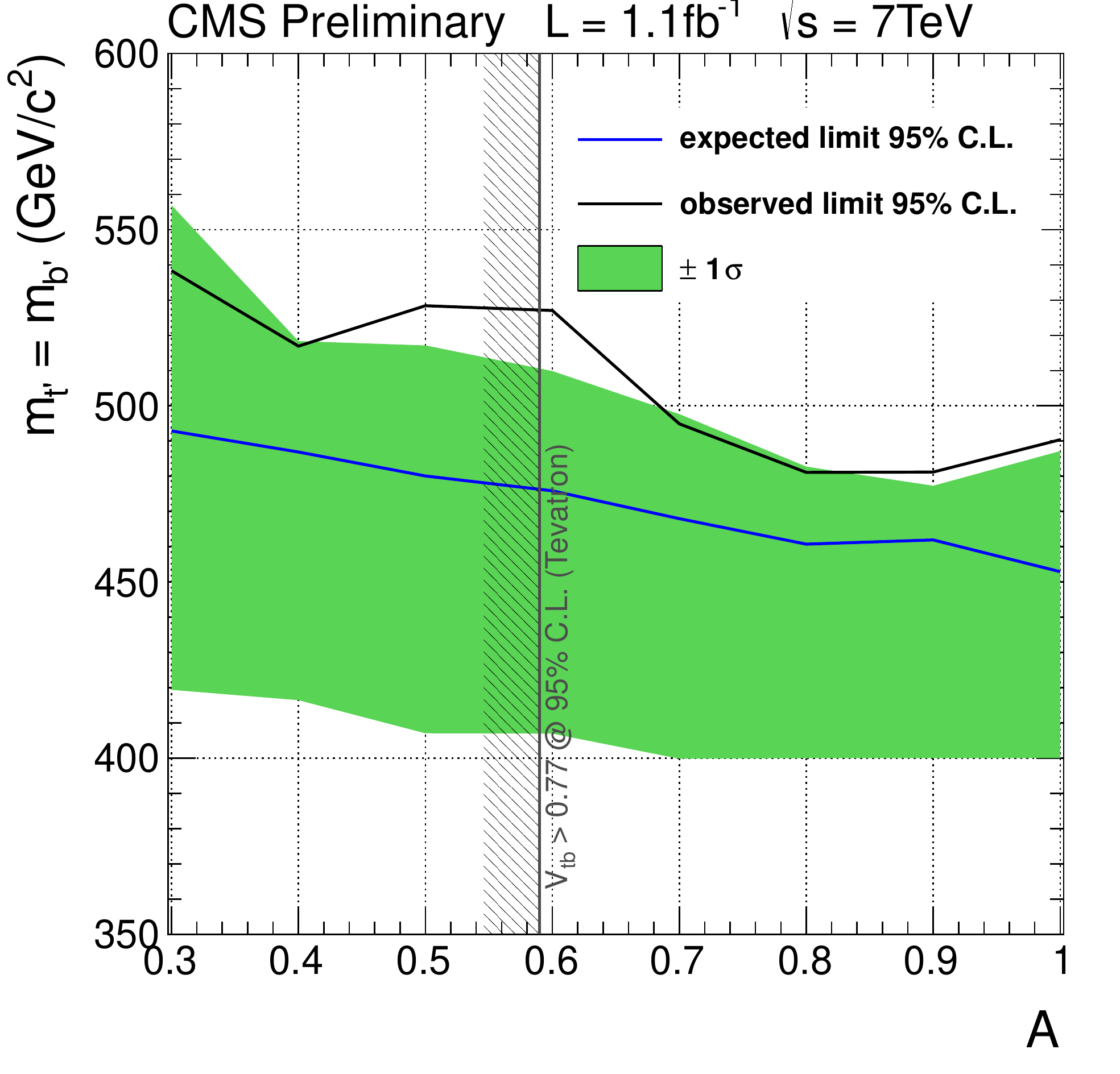} }
\caption{Expected and observed exclusion limits for $M_{t'} = M_{b'}$ as a function of the 
four-generation CKM parameter A as described in the text.  The parameter values below the 
solid black line are excluded at 95\% C.L.}
\label{fig:q4thgen}       
\end{center}
\end{figure}

We additionally consider mixing between the third and fourth generations and 
search for fourth generation quarks produced singly ($t'b$ or $b't$) 
or in pairs as a function of the four-generation CKM matrix parameter $A$, where  
$V_{tb} = V_{t'b'} = \sqrt{A}$ and $V_{t'b} = V_{tb'} = \sqrt{1-A}$~\cite{4thgenCKM}.  
This inclusive search requires at least one of the $W$ bosons in the decay chain to 
decay to a muon and neutrino, and events are categorized according to 
the number of $b$-tagged hadronic jets and reconstructed 
$W \rightarrow q\bar{q}'$ candidates.  We find no evidence for fourth generation quarks
in this search.  Assuming degenerate $t'$ and $b'$ masses, we summarize the 95\% C.L. 
exclusion limits on fourth generation quark masses as a function of $A$ in 
Figure~\ref{fig:q4thgen}.
 
A search for vector-like quarks ($T$) with charge $|q| = \frac{2}{3}$ is also performed 
by CMS~\cite{vecTop} using 1.1~fb$^{-1}$ of collision data.  We search for $pp \rightarrow TT$ production, 
followed by $T \rightarrow tZ \rightarrow bWZ$.  Motivated by this decay chain, 
we search for events with at least two jets and at least three leptons, requiring that two of the leptons
be consistent with $Z \rightarrow \ell \ell$ decay.  Based on these requirements, 
backgrounds are dominated by events with real $Z$ bosons.  We find no excess beyond SM 
expectations, and conclude that $M_T > 475$~GeV at the 95\% confidence level.

\section{Search for Extra Dimensions}

The great disparity between the Planck scale ($M_{\rm Pl} \sim 10^{19}$~GeV) 
and 
the electroweak scale ($M_{\rm EW} \sim 10^3$~GeV) is a great mystery of the Standard Model.  
The model proposed by Arkani-Hamed, Dimopoulos, and Dvali (ADD) includes a framework for 
addressing this hierarchy problem.  Taking the fundamental scale to be of the same order 
as the electroweak scale ($M_D \sim M_{\rm EW}$), the observed weakness of gravitational 
interactions (equivalently, the large Planck mass) would be a consequence of the universe 
having compactified extra dimensions.
In the ADD model, the SM gauge interactions are constrained to the familiar 
3 (space) + 1 (time) dimensional subspace while 
gravity is allowed to propagate through the entire multidimensional space.  Taking these 
hypothesized extra dimensions into account, the 
Planck scale is related to the fundamental scale by $M_{\rm Pl}^2 \sim M_D^{n+2} r^n$, 
where $r$ and $n$ are the size and number of extra dimensions, respectively.  
If $M_D$ is at the TeV scale, it would be possible to produce gravitons at the LHC.

\begin{figure}[htb]
\begin{center}
\resizebox{0.85\columnwidth}{!}{%
\includegraphics{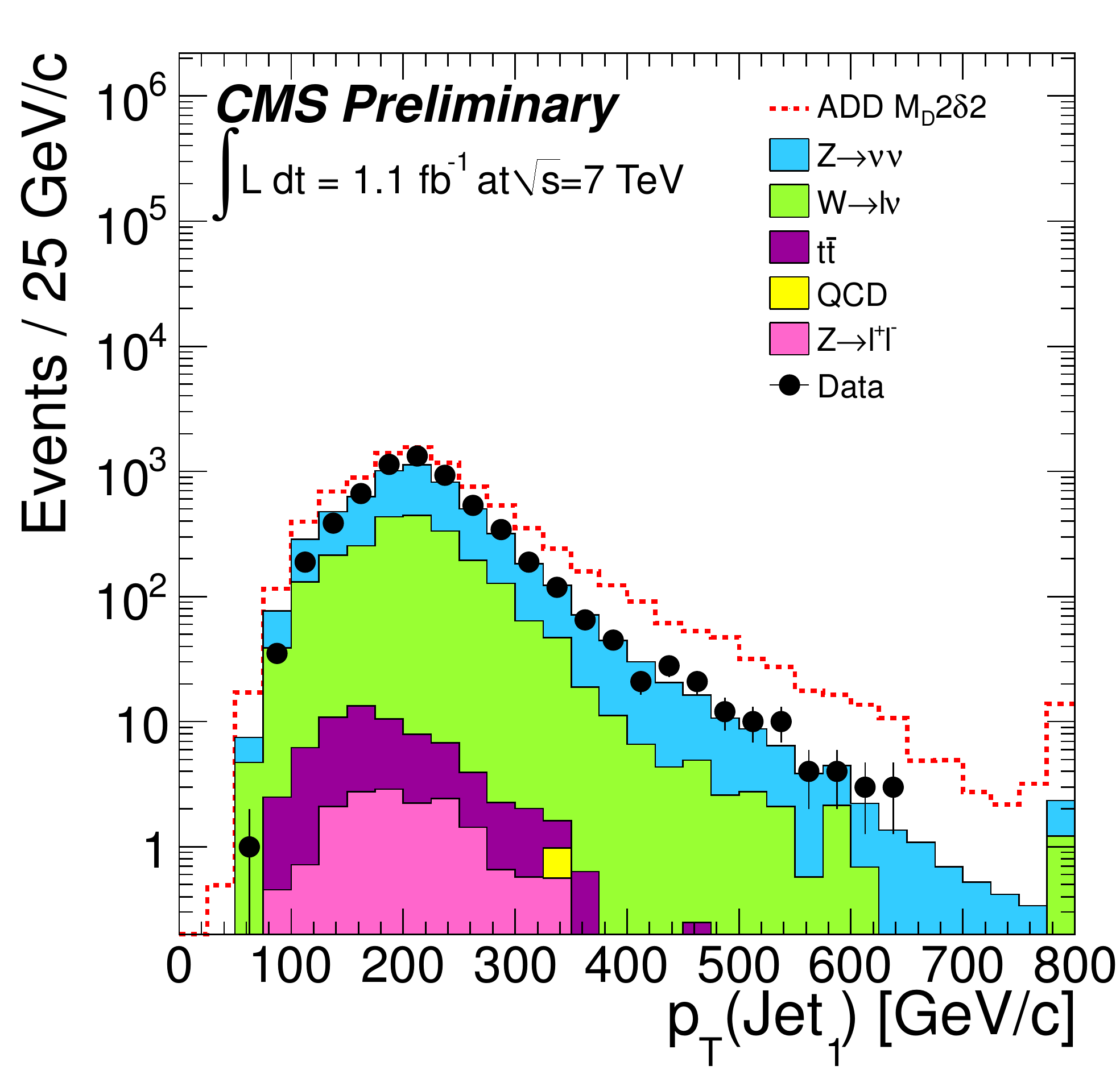} }
\caption{Transverse momentum distribution for the leading jet in the jet+$E_T^{miss}$ search for extra dimensions.
The leading jet is required to have $p_T > 100$~GeV and be reconstructed within the acceptance of the 
CMS tracker.}
\label{fig:monojet}       
\end{center}
\end{figure}

Gravitons ($G$) could be produced directly at the LHC via the processes $gg \rightarrow qG$, 
$q\bar{q} \rightarrow gG$, $qg \rightarrow qG$, and $q\bar{q} \rightarrow gG$.  
As the graviton escapes detection, we must search for evidence for extra dimensions in 
final states that include large $E_{T}^{\rm miss}$.  CMS performs two searches 
of this nature, as we search for final states involving a real graviton balanced by 
a photon~\cite{monophoton} or hadronic jet ($j$)~\cite{monojet}.
In each case, events are rejected if additional jets or isolated tracks are found with 
significant transverse energy, and additional requirements suppress beam backgrounds, cosmic 
ray signatures, and instrumental noise that may mimic the desired final state. 
We present the leading jet transverse momentum spectrum for the $j + E_{T}^{\rm miss}$ 
channel in 
Figure~\ref{fig:monojet}.
Both the $\gamma + E_{T}^{\rm miss}$ and $j + E_{T}^{\rm miss}$ searches find no 
evidence for direct graviton production using 1.1~fb$^{-1}$ of data, and set limits on 
the fundamental mass scale $M_D$ that depend on the number of extra dimensions 
considered.

Virtual graviton exchange is also possible at the LHC, which could lead to 
enhanced non-resonant dimuon production, which we consider in~\cite{dimuonadd}.  
Using 1.1~fb$^{-1}$ of data, we find results are consistent with SM expectations.  We 
consequently set 95\% C.L. lower limits on the fundamental mass scale depend on the 
number of extra dimensions and additional theoretical conventions used to describe virtual 
graviton production.

\subsection{Search for Microscopic Black Holes}

One of the exciting potential consequences of theoretical models with extra spatial 
dimensions and lowscale quantum gravity is the copious production of microscopic black holes 
in particle collisions at the LHC.  We analyze events with large total transverse energy, 
searching for multiple high energy jets, leptons, and photons which are expected from the 
decay of black holes produced at LHC energies~\cite{bh}.
Backgrounds from Standard Model
processes, which are dominated by QCD multijet production, are in good agreement with 
the results found in data.  We therefore set limits on the minimum black hole mass that 
depend on the number of ADD extra dimensions and the fundamental mass scale $M_D$.

\section{Search for R-Parity Violation in Multilepton Events}

One common feature of many Supersymmetric (SUSY) extensions to the Standard Model 
is conservation of ``R-parity''.  Conserving this quantity forces Supersymmetric particles 
to be produced in pairs and ensures the lightest SUSY particle (LSP) is stable and thus a  
good dark matter candidate.  As the LSP escapes detection, many SUSY searches require $E_T^{\rm miss}$ 
in the final state to account for the loss of the LSP.

\begin{figure}[htb]
\begin{center}
\resizebox{0.85\columnwidth}{!}{%
\includegraphics{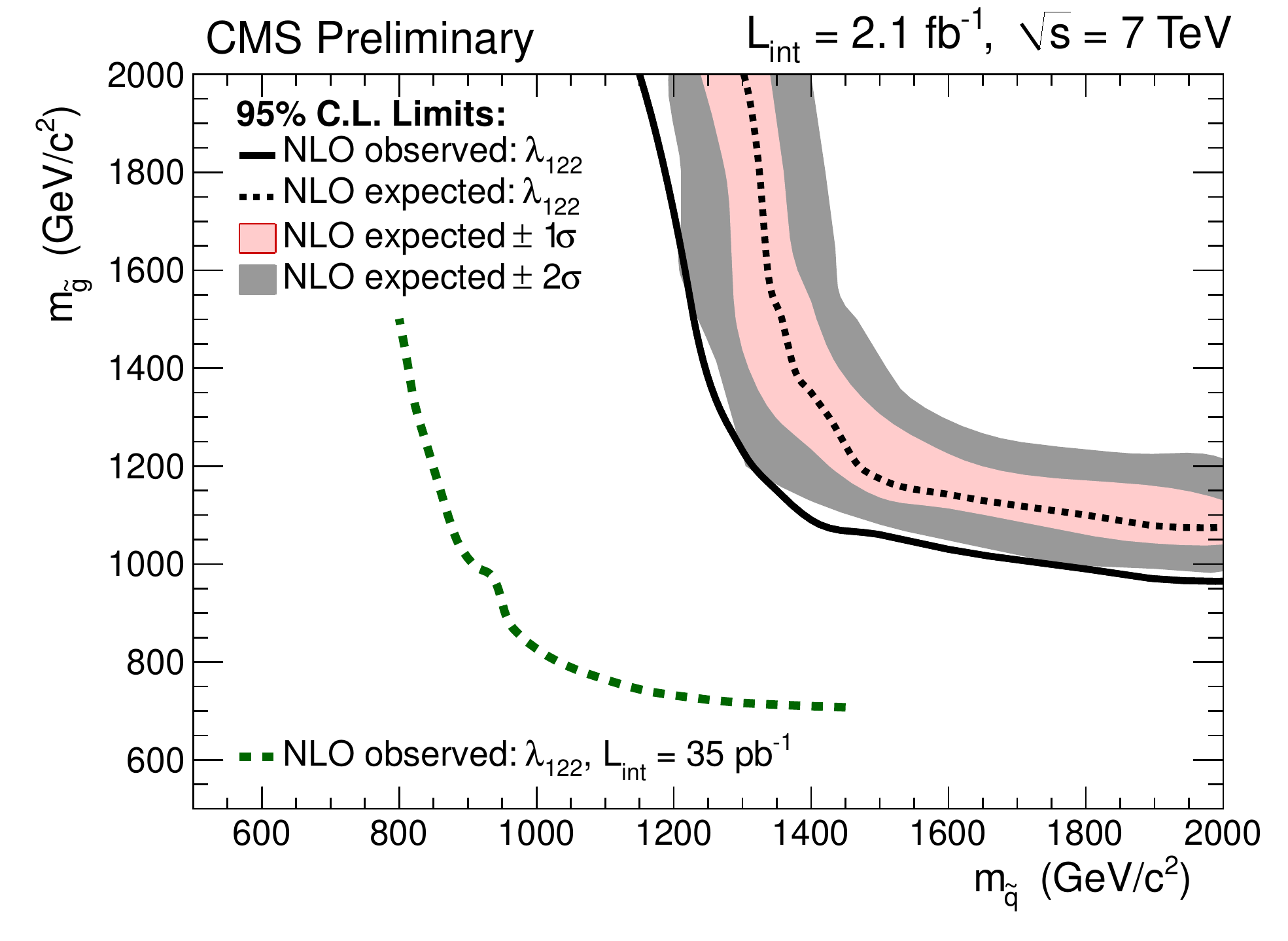} }
\caption{95\% C.L. limits for R-Parity violating coupling $\lambda_{122} = \lambda_{e\mu\mu}$ as
a function of the squark and gluino masses for a SUSY topology described in~\cite{rpv}. The
observed limits, along with limits expected in the absence of signal are shown, in addition to the
uncertainty in the expectation. Masses to the left of the curves are excluded.}
\label{fig:rpv}       
\end{center}
\end{figure}

If we instead consider models with R-parity violating interactions, the LSP loses stability 
and consequently decays to SM particles that could be observed in the detector.  As 
$E_T^{\rm miss}$ loses significance as a search quantity, we instead 
search for deviations from Standard Model expectations in events with significant total 
transverse energy~\cite{rpv}.  Using 2.1~fb$^{-1}$ of data collected in 2011, we consider 
52 exclusive channels, categorized according to the number of leptons ($\ell = e, \mu$) 
and same-flavor $\ell^+ \ell^-$ pairs in the event, the number of reconstructed $\tau$ leptons 
in the event, whether or not a $Z \rightarrow \ell \ell$ decay is reconstructed, and the 
total transverse energy in the event.  Several channels are dominated by SM processes 
and serve as control channels, while other channels expect negligible SM background contributions.
Data-driven methods are used to estimate background levels, including an estimate of 
asymmetric photon conversions using $Z \rightarrow \mu \mu \gamma \rightarrow 4 \mu$ events.
We see good overall agreement with background expectations across all channels, 
and set limits on coupling strengths for R-parity violating SUSY interactions.  One such 
limit is shown in Figure~\ref{fig:rpv}.

In two channels with high total transverse energy, four leptons, and at least one same-flavor $\ell^+ \ell^-$ pair,
we expect low backgrounds from SM processes but find a total of two events, finding one event 
each in channel (with or without a $Z$ candidate).  While interesting, additional luminosity is 
required before we can access the significance of the result.

\section{Search for Heavy Long-Lived Particles}

\subsection{Charged Particles}

Heavy quasi-stable charged particles appear in various extensions to the
Standard Model.  Heavy long-lived particles with a hadronic character, such as 
gluinos or stop squarks, will hadronize in flight and form meta-stable bound states with 
quarks and gluons.  If the lifetime of these so-called ``R-hadrons'' produced at the LHC exceeds
a few nanoseconds, the particles will travel over distances that are comparable
or larger than the size of the CMS detector.  Depending on particle lifetime and 
interaction with the detector material, CMS applies two complimentary techniques when 
searching for heavy long-lived charged particles.

The first approach assumes that a significant fraction of 
the R-hadrons will have a velocity smaller than $0.9 c$.  As a result, these particles 
will be directly observable through the distinctive signature 
of a high momentum particle with an anomalously large rate of ionization energy loss 
in the CMS tracker and an anomalously long time-of-flight as measured by the CMS 
Muon system~\cite{hscp}.  We search for R-hadrons traversing the entire detector using both 
the tracker and muon systems, and additionally search using only the tracker information to 
account for the possibility that the particle may become neutral before reaching the muon 
detectors.  Using 1.1~fb$^{-1}$ of data collected in 2011, searches for anomalous charged 
tracks allow us to exclude the production of gluinos, hadronizing into stable R-gluonballs with 
10\% (50\%) probability, with masses below 899 (839)~GeV at 95\% confidence level.  The 
analogous limit on stop production is 620~GeV.

We also perform a search for low velocity ($v < 0.4 c$) R-hadrons, where the energy loss 
is great enough that a significant fraction of the produced particles come to rest in the 
CMS detector volume~\cite{stopped}.  These ``stopped'' R-hadrons will then decay to 
hadronic jets at some later time depending on the unknown particle lifetime.  The decay of 
these particles will occur out-of-time with respect to LHC $pp$ collisions, so our 
online (trigger) selection for stopped particles places a veto on the beam presence.  In a
dataset with a peak instantaneous luminosity of $1.3 \times 10^{33}$~cm$^{-2}$~s$^{-1}$, 
an integrated luminosity of up to 886~pb$^{-1}$ lifetime, and a search interval corresponding 
to 168 hours of trigger live time, no significant excess above background (mostly 
from instrumental noise) was observed.  Limits at the 95\% confidence level on R-hadron pair 
production extend over 13 orders of magnitude of gluino lifetime.  

\subsection{Neutral Particles}

\begin{figure}[htb]
\begin{center}
\resizebox{0.47\columnwidth}{!}{%
\includegraphics{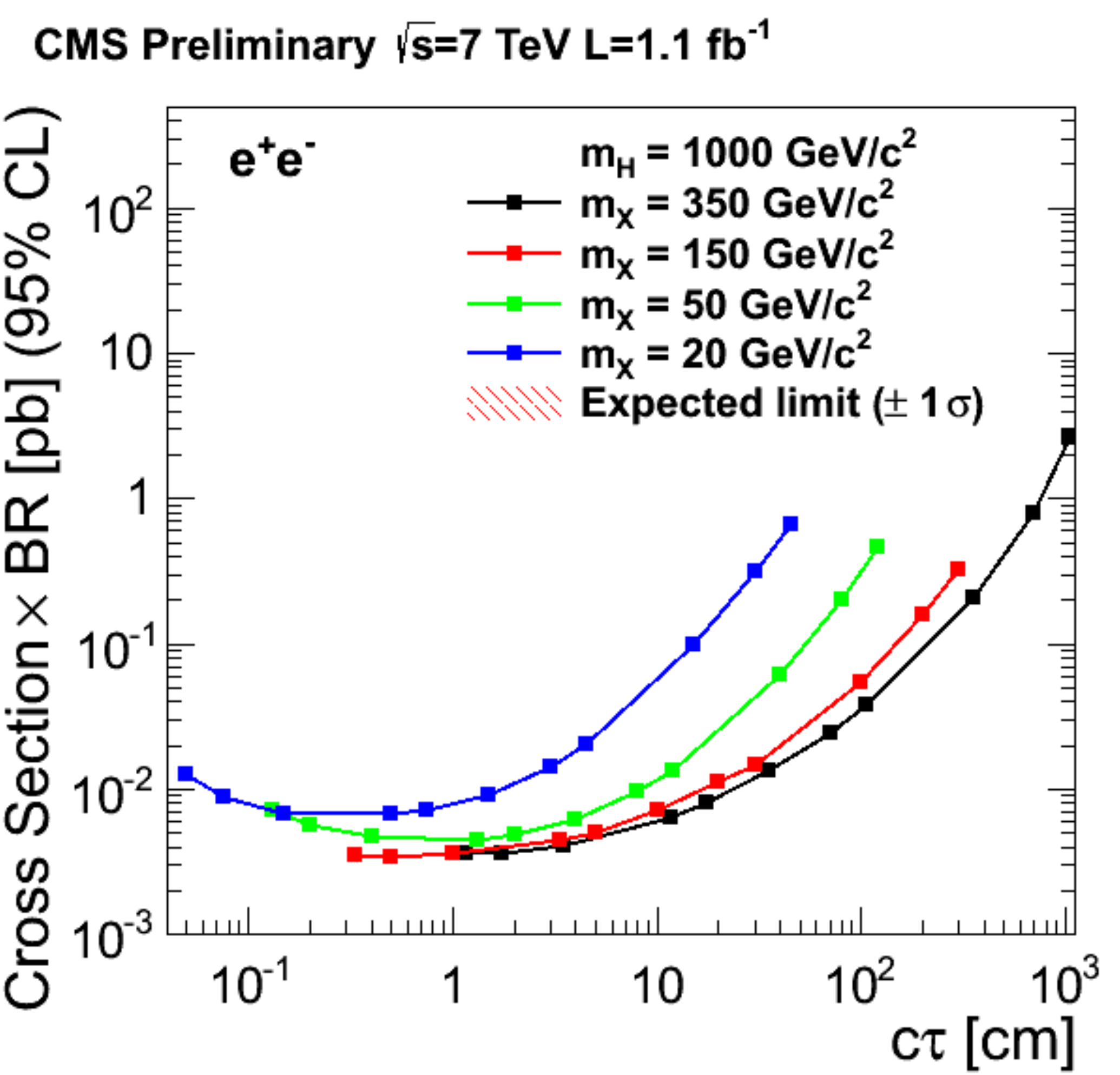} }
\resizebox{0.47\columnwidth}{!}{%
\includegraphics{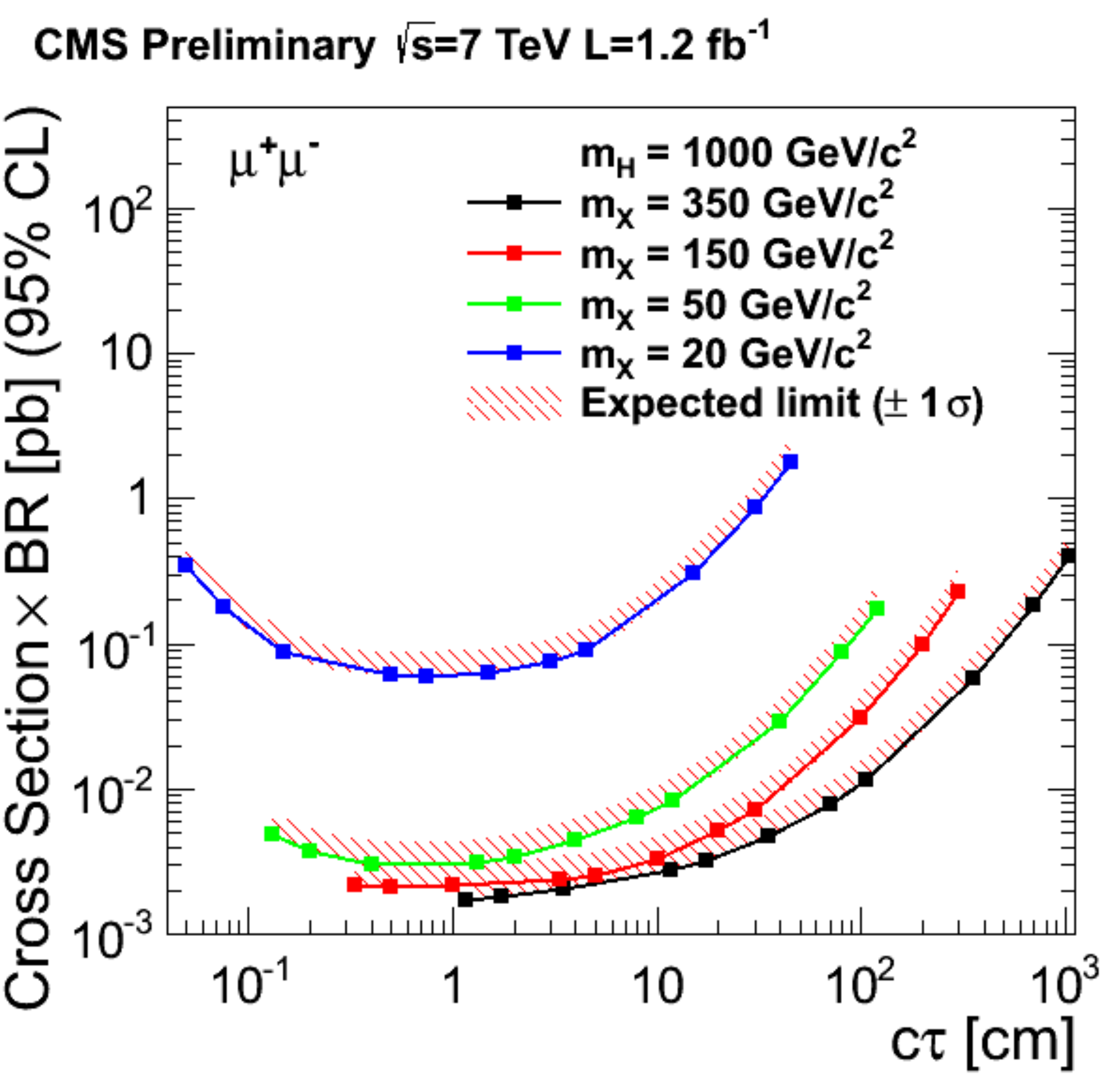} }
\caption{95\% confidence level upper limits on the cross section and branching fraction for 
the assumed $H \rightarrow \chi \chi$ decay, with $\chi \rightarrow ee$ (left) and $\chi \rightarrow \mu \mu$ (right), 
assuming a Higgs mass of 1000~GeV. The expected limit band for the electron channel is too small to be
visible.}
\label{fig:dilepLL}  
\end{center}
\end{figure}

CMS also searches for long-lived neutral particles that are present in several exotic 
physics models.  In one such search, we consider pair production of long-lived neutral 
particles (e.g. $H \rightarrow \chi \chi$), each of which subsequently decays to a dielectron 
or dimuon pair~\cite{dilepLL}.  
As the massive neutral particles travel some distance in the detector before decaying, 
the $\chi \rightarrow \ell^+ \ell^-$ decay vertex will be displaced with respect to the 
$pp$ collision vertex.  Using 1.1~fb$^{-1}$ of collision data, we select $e^+ e^-$ and $\mu^+ \mu^-$ 
combinations that have large transverse decay lengths relative to the decay length uncertainty. 
A single event survives the $e^+ e^-$ selection, consistent with expectations from $Z \rightarrow ee$
decay.  No events survive the muon selection.  In Figure~\ref{fig:dilepLL} we present the 95\% C.L. exclusion limits 
as a function of the decay length of the heavy neutral particle.

In some cases, the long-lived neutral particle may decay exclusively to other neutral particles 
(one example is the decay of a neutralino to a photon and a gravatino).  In these cases, we 
can exploit the large amount of tracker material in CMS to estimate the particle 
lifetime~\cite{pointing}.
We search for photons that convert in the detector to a dielectron pair.  As the four-momentum 
of the $\gamma \rightarrow e^+ e^-$ pair points along the same direction as the parent photon, 
the tracks of the electrons can be precisely reconstructed and used to calculate the photon trajectory and, 
in particular, the impact parameter of the photon with respect to the interaction point.
Examining 2.1~fb$^{-1}$ of collision data for a photon with significant impact parameter 
in association with missing transverse energy (due to the escaping gravatino), we find the 
data is in good agreement with expectations from SM processes and determine the upper limits for
the cross section for pair-production of neutralinos, each of which decays into one photon and invisible
particles, as a function of neutralino lifetime.

\section*{Conclusion}

In this Note we summarize several searches for exotic new physics processes not 
currently explained by the Standard Model.  In each search, the data are 
in good agreement with expectations from Standard Model processes.  We proceed to 
set 95\% C.L. limits on the parameters and particle masses relevant for various 
new physics scenarios.

\section*{Acknowledgments}

The author wishes to thank the organizers of the Hadron Collider Physics Symposium for 
the rich and interesting physics program and the excellent conference location.

\end{document}